\documentclass[lettersize,journal]{IEEEtran}
\usepackage{amsmath,amsfonts}
\usepackage{booktabs} 
\usepackage{siunitx} 
\usepackage{amssymb}  
\usepackage{pifont}   
\usepackage{booktabs}
\usepackage{array}
\usepackage[caption=false,font=normalsize,labelfont=sf,textfont=sf]{subfig}
\usepackage{listings}
\usepackage{textcomp}
\usepackage{stfloats}
\usepackage[ruled,linesnumbered,vlined]{algorithm2e}
\usepackage{url}
\usepackage{verbatim}
\usepackage{graphicx}
\usepackage{cite}
\usepackage[backref]{hyperref}
\usepackage{amssymb}
\usepackage{enumitem}
\usepackage{color}
\usepackage{multirow} 
\usepackage{makecell} 
\usepackage{booktabs} 
\usepackage{pdflscape} 

\DeclareRobustCommand*{\IEEEauthorrefmark}[1]{%
    \raisebox{0pt}[0pt][0pt]{\textsuperscript{\footnotesize\ensuremath{#1}}}}

\makeatletter

\newcommand{\Rmnum}[1]{\expandafter\@slowromancap\romannumeral #1@}

\newcommand{\cmark}{\ding{51}}  
\newcommand{\xmark}{\ding{55}}  

\makeatother
\RequirePackage{float,graphicx}

\SetCommentSty{mycommfont}

\SetKwInput{KwInput}{Input}                
\SetKwInput{KwOutput}{Output}              
\begin{document}

\title{SAFLA: Semantic-aware Full Lifecycle Assurance Designed for Intent-Driven Networks}
\author{Shiwen Kou, Chungang Yang, Mingji Wu
\author{
\IEEEauthorblockN{
Shiwen Kou\IEEEauthorrefmark{1},
Chungang Yang\IEEEauthorrefmark{1},
Mingji Wu\IEEEauthorrefmark{1},
}

\IEEEauthorblockA{\IEEEauthorrefmark{1}The State Key Laboratory on Integrated Services Networks, Xidian University, Xi'an, China.}

\IEEEauthorblockA{shiwenkou@gmail.com, chgyang2010@163.com; 15120292580@163.com}

\IEEEauthorblockA{Corresponding Author: Shiwen Kou \quad Email: shiwenkou@gmail.com}}


\thanks{This paper was produced by the IEEE Publication Technology Group. They are in Piscataway, NJ.}
\thanks{Manuscript received April 19, 2021; revised August 16, 2021.}}

\markboth{Journal of \LaTeX\ Class Files,~Vol.~14, No.~8, August~2021}%
{Shell \MakeLowercase{\textit{et al.}}: A Sample Article Using IEEEtran.cls for IEEE Journals}


\maketitle

\begin{abstract}

Intent-driven Networks (IDNs) are crucial in enhancing network management efficiency by enabling the translation of high-level intents into executable configurations via a top-down approach. The escalating complexity of network architectures, however, has led to a semantic gap between these intents and their actual configurations, posing significant challenges to the accuracy and reliability of IDNs. While existing methodologies attempt to address this gap through a bottom-up analysis of network metadata, they often fall short, focusing primarily on intent extraction or reasoning without fully leveraging insights to tackle the inherent challenges of IDNs. To mitigate this, we introduce SAFLA, a semantic-aware framework specifically designed to assure the full lifecycle of intents within IDNs. By seamlessly integrating top-down and bottom-up approaches, SAFLA not only provides comprehensive intent assurance but also effectively bridges the semantic gap. This integration facilitates a self-healing mechanism, substantially reducing the need for manual intervention even in dynamically changing network environments.  Experimental results demonstrate the framework's feasibility and efficiency, confirming its capacity to quickly adapt intents in response to network changes, thus marking an important advancement in the field of IDNs.

\end{abstract}


\section{Introduction}

In today's world, digital connectivity is integral to virtually every aspect of our lives. However, the conventional methods of managing network service relies on static and script-based systems, which may not always meet the diverse and changing requirements of user who seek more adaptable services. To tackle this challenge, Software-Defined Networking (SDN) steps in, redefining the behavior of data plane switches through a software-centric approach. Nevertheless, a significant gap persists between the intricate complexities of network management and the dynamic needs of users. This growing divergence underscores the critical need for reducing human intervention in the control loop. In response, Intent-Driven Networks (IDNs) have been introduced as an innovative solution, offering automation for more adaptive and responsive network management\cite{Pang}. As a self-orchestrating network, IDNs permit users to specify their service needs through high-level intents\cite{Zeydan}, obviating the need to worry about intricate details of how these requirements are translated into hardware configurations. Previous research indicated that IDNs predominantly utilize a top-down methodology, primarily centering on an intent refinement process, which is geared towards translating high-level human languages into intent policies\cite{ouyang}. In the top-down methodology, the process of intent refinement often overlooks the negotiation of underlying states, specifically configuration and network statuses. This oversight presents challenges in ensuring the intent's feasibility. In summary, the major novel technical challenges of the top-down method are as follows:

\begin{itemize}
\item \textbf{Completeness:}  We observe that network configurations and network states serve as excellent examples of network knowledge, which illustrate its internal logic and operations. However, this knowledge has been largely overlooked in prior studies.
\item \textbf{Correctness:} As the network state undergoes dynamic changes, deployed intents may experience disruption if significant alterations do not align with the requirements of the user's intents. An intent mismatch detection mechanism is required to maintain the intent's full lifecycle.
\item \textbf{Autonomy:} It is crucial to recover intent when disruption emerges due to network changes, enabling automatic maintenance of semantic alignment between the user's intent and the real configurations.
\end{itemize}

Based on the aforementioned analysis, we notice that the top-down approach in recent IDNs often encounters limitations due to its partial visibility to fully comprehend and integrate underlying knowledge, indicating a notable semantic gap. This gap suggests that there is an urgent need for frameworks within IDNs that can adeptly navigate and resolve intent disruptions from both a comprehensive (top-down) and a detailed (bottom-up) perspective.

To address the semantic gap, a few methods employing a bottom-up approach have been introduced. These methods involve an in-depth analysis of network operations, with a particular focus on the detailed aspects of network hardware and software status. Among these, PROVINTENT \cite{provintent} introduced an intent provenance model that is critical in tracking and interpreting the state and semantics of network intents, thereby providing key insights into the operational dynamics of network policies. Similarly, SCRIBE\cite{scribe} adopted a deterministic method to reverse-engineer configuration information from configuration files, ultimately presenting it in the Nile language, which simplifies the understanding and management of network configurations. Furthermore, the Policy Intent Inference (PII) system underscores that the semantic gap between higher and lower network layers is a major barrier in deploying intent-driven approaches in traditional networks\cite{pii}. PII introduced a system to infer policy intents, effectively addressing this challenge. Collectively, these methods concentrate on the semantic flow of software or hardware status, addressing the network visibility issues mainly in a bottom-up paradigm. 

\begin{figure}
\centering
\includegraphics[width=3in]{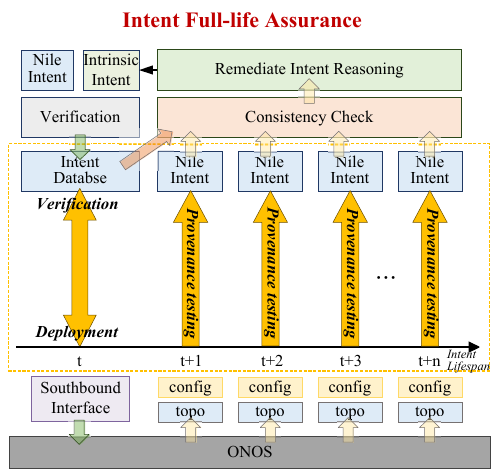}
\caption{The motivation of the proposed framework.}
\label{motivation}
\end{figure}

While these methods effectively reduce the semantic gap and provide valuable insights into the network state, there still exists a pressing need for an integrated framework that adeptly combines both approaches, leaving the stated challenges for the current IDNs framework unsolved. Therefore, we propose the Semantic-aware Framework for Full Lifecycle Assurance (SAFLA), which aims to reconceptualize the IDNs process by synergizing both perspectives. In contrast to existing methods, the motivation behind our proposed framework, as depicted in \ref{motivation}, centers on ensuring the full lifecycle integrity of network intents. By systematically analyzing network configurations from the bottom-up and aligning these with the user's intent from the top-down, our approach aims to guarantee the correctness of intents throughout their entire lifecycle. This dual perspective allows for a comprehensive understanding and management of network behaviors, ensuring that operational configurations accurately reflect the strategic objectives set forth by network administrators. In summary, the main contributions of our work are listed as follows:

%
%
%
\begin{itemize}
\item To the best of our knowledge, such a holistic IDNs framework that bridges both top-down and bottom-up methodology does not yet exist in the current IDNs landscape. Therefore, we present an integrated framework for IDNs to ensure comprehensive lifecycle assurance of network intents by analyzing the underlying knowledge. As a result, SAFLA enhances the robustness and adaptability of network management by autonomously maintaining the alignment between user intents and network configurations. We believe this paper could provide novel insight into IDNs, preserving intent within its full-life cycle.
\item SAFLA can detect intent inconsistency and repair such errors due to network dynamics, which helps reduce the complexity of manual troubleshooting and ensures network operations remain aligned with the user's intent.
\item We present the results of experiments on routing configurations in SDN. SAFLA detects intent inconsistency and counteracts it in a matter of a few seconds, showcasing its efficiency and feasibility in real-time network management.
\end{itemize}
\section{Related Works}
\subsection{Top-down Methodology for IDN}


Currently, IDN research follows two technical approaches: top-down and bottom-up paradigms. The top-down approach aims to translate users' uttered intent expectations into low-level network policies and configurations. Extensive studies have proposed natural language processing methods to translate ambiguous human intent into concrete network configurations. Several studies have applied natural language processing (NLP) to parse intents and generate policies. For example, the authors in \cite{ouyang} present an intelligent intent translation framework. It employs a bi-directional long and short-term memory and conditional random field. Other works like VIVoNet\cite{vivonet} supply a voice-assistant interface for operators to input intents and convert intents to network configurations afterward. Like other deep learning methods, this data-driven approach enable predictive translation but require large training data. To improve translation performance, some works explore intermediate representations between intents and policies. The authors in \cite{refinement} proposed an intermediate intent language called Network Intent LanguagE (Nile) to address challenges in decoupling policy extraction from deployment. Building on this, Lumi\cite{lumi} incorporates users' real-time feedback into the intent confirmation state. By this approach, Lumi's design leverages user expertise as a critical resource for constantly augmenting and updating its existing dataset, which in turn, significantly bolsters the accuracy of the intent translation. Recently, SMART\cite{smart}, ingeniously introduced the concept of an intent quintuple, denoted as $\langle \text{domain, attribute, object, operation, result} \rangle$, which specified the scope that an intent expression should cover. This quintuple serves as a comprehensive framework that significantly enhances the precision and efficacy of network management. 
Despite the success of the aforementioned efforts, the top-down paradigm may fail when deploying new intents with limited network capabilities. As such, the top-down approach remains limited in generalized intent translation and policy enforcement. This motivates research into bottom-up paradigms, which will be discussed next.

\subsection{Bottom-up Methodology for IDN}

To address the limitation of the top-down methodology in dynamic network management and intent assurance, some studies have explored a bottom-up approach. The bottom-up approach collects underlying network configurations and intrinsic network knowledge to infer the intents behind existing policies. 
For example, SCRIBE introduces a bottom-up approach to extract higher-level intent from the lower-level configuration files. As a variant, PII improves upon it by incorporating semantic metadata in the network as auxiliary information to aid in extracting already deployed intents. Moreover, PROVINTENT extends the SDN control plane by recording the provenance and evolution of all intents, and reasoning about intents by analyzing past intent state changes. A key advantage of the bottom-up approach is that it strengthens the relations between the current configurations and past intents, providing a heuristic to derive applicable and sound policies. While existing IDNs approaches effectively employ either a top-down or bottom-up methodology, they often fall short of merging these strategies in a way that ensures semantic consistency across the full lifecycle of intents. This shortfall points to the necessity for an IDNs framework that not only combines the strengths of both methodologies but also guarantees the semantic integrity of intent throughout their entire lifecycle. Such a framework is crucial for achieving a deep and dynamic alignment of intent semantics with the ever-evolving requirements of the network, ensuring that intents are consistently understood and applied as intended.

\section{An Overview of the Proposed Framework}
In this section, we introduce the proposed SAFLA and detail its integration with the existing top-down methodology through a deterministic bottom-up approach. Moreover, we describe our deterministic method for inferring deployed intents. Our framework proactively responds to network dynamics by automatically reasoning and deploying remediation intents based on these inferred results.

\begin{figure*}
\centering
\includegraphics[width=5in]{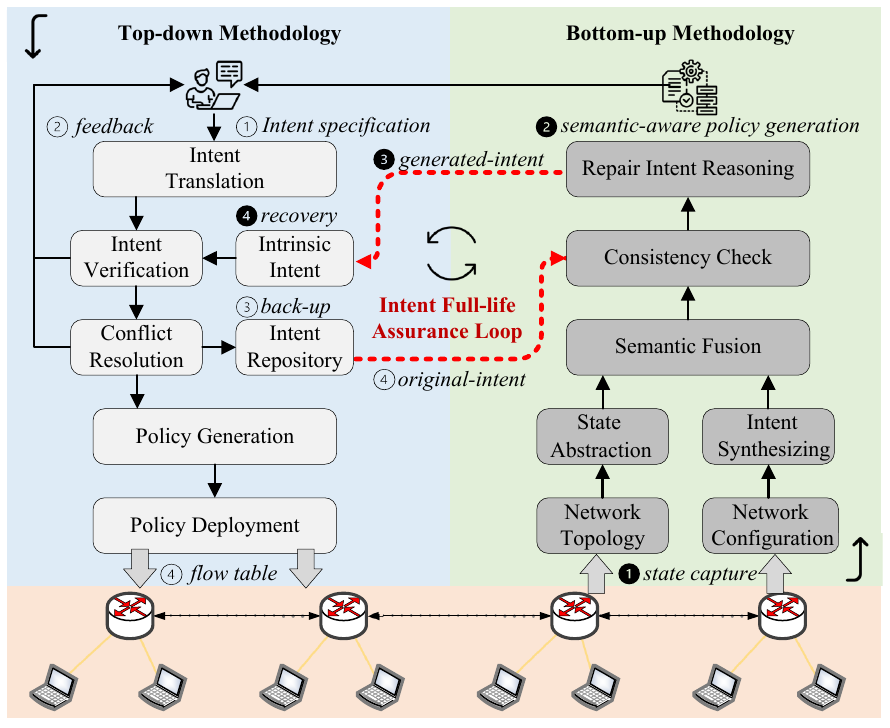}
\caption{The architecture of the semantic-aware full lifecycle assurance framework}
\label{fig:picture001}
\end{figure*}

The proposed framework is built upon three fundamental components: top-down intent refinement, bottom-up intent extraction, and repair intent reasoning. As depicted in Fig. \ref{fig:picture001}, SAFLA takes both the network configurations and the underlying network topology as its input. Through semantic feature extraction, it discerns the remediation intent embedded within the configurations and states. This feature is crucial, especially when the network's current capabilities conflict with the user's original intent due to changes in the network. The configuration analyzing component extracts rules from the network policies. The network abstracting component monitors the network status in real time. Lastly, the remediate intent reasoning component pinpoints the remediation intent from the derived semantic features. In the following sections, we will delve into each of these components in greater detail.




\subsection{Top-down Components in Intent-Driven Networks}
In IDN, a top-down approach plays a pivotal role in steering network behaviors and operations towards desired outcomes. The process begins with \textit{Intent Refinement}, a crucial phase where high-level business objectives or user intents are translated into detailed, deployable network policies. 
This refinement involves decomposing broad, often abstract goals into specific, measurable network configurations. Following this, \textit{Intent Verification} is employed to ensure that the refined intents align with the overall network capabilities and constraints. This step acts as a safeguard against impractical or conflicting intents, incorporating simulations or predictive models to verify the feasibility and efficiency of the proposed network configurations. Lastly, \textit{Conflict Resolution} is an essential mechanism within IDN to address any discrepancies or clashes that arise between multiple intents or between intents and existing network policies. This involves a sophisticated analysis to identify and reconcile conflicts, ensuring harmonious coexistence of various intents and maintaining network integrity. Through these integrated processes, the top-down approach in IDN fosters a seamless translation of user objectives into efficient network operations, paving the way for agile and responsive network environments. For further details on the intricacies and implementation of these processes within the Intent-Driven Networks framework, we refer the interested reader to our previously published literature\cite{ouyang,kid,song,ontology}.

\subsection{Bottom-up Analysis and Intent Assurance}
As depicted in Fig. \ref{fig:picture001}, our methodology primarily relies on configuration files and the current network status as inputs. These configuration files are exported from the underlying switches, whereas the network state includes elements such as endpoint names, links, interfaces and IP addresses.
In this section, we outline a bottom-up methodology for analyzing network behavior and translating it into an intent-level language. Our methodology takes flow tables exported from SDN switches as input. Upon acquiring these flow tables, during step 1, we categorize them into two primary groups: forwarding and functional. Based on their specific functionalities, we consolidate them in a centralized database, synthesize the data, and then represent them using an intermediate format. In step 2, we identify network entities to produce an abstract model. Step 3 involves augmenting this model with supplementary data sourced from NSKG, culminating in the Aggregated Model. This model offers an enhanced depiction of switch behavior, characterized as meta-intents. Finally, in step 4, we convert these meta-intents into the network intent language, where we carry out conflict evaluation and policy management.
\subsubsection{State Abstraction}
\label{monitoring}
Given the increasing complexity and diversity of modern networks, our framework initiates the process by analyzing the extracted network state. To tackle the challenges of network heterogeneity, we employ a knowledge graph (KG) to abstract the physical network. This KG enables a comprehensive network analysis across various dimensions, levels, and granularities, allowing for an in-depth exploration of network attributes from different perspectives. Furthermore, we use a vendor-neutral template to extract essential features from the network state, including topology, endpoint groups, links, and their associated attributes. In our case, we focus on endpoint entities and their related attributes. It is worth noting that these network endpoints largely remain static, prompting us to gather it using both real-time and offline methods. As we collect this data, our framework automatically constructs a Network State Knowledge Graph (NSKG) from the relevant entities and attributes. Consequently, this NSKG can dynamically update its nodes and relation properties based on real-time data. This provides a way to represent dynamic network changes in a format that's both machine-compatible and human-readable. The NSKG is capable of not just detecting changes in network status, but also of being utilized to determine whether these changes could result in an intent disruption. If such disruption occurs, the NSKG can provide data to infer the underlying remediation intent and act on it by utilizing the attributes of the affected endpoint group.

\label{analyzing}

\subsubsection{Intent Synthesising}

We first export flow tables from the programmable SDN switches. Our objective is to synthesize these flow tables into a standard intermediate representation that denotes high-level intents. It is crucial to ensure that the exported flow tables are complete and unaltered as they form the most primitive description of the SDN switches. Our ultimate goal is to unify the underlying flow entries, which will abstract the underlying configuration files into an intermediate representation. By synthesizing such a representation, it became feasible to analyze and extract high-level intents. 

In SDN environments, the flow tables play a critical role in directing how a switch handles traffic. A flow table consists of multiple flow entries, each of which specifies the rules for handling particular traffic patterns. the flow entries have match fields that determine which packets align with the entry and actions that define what should happen to the matching packets, such as forwarding or dropping them. Moreover, the entries have priority attributes that decide which entry should be applied when multiple matches are found for a packet. Such flow tables and their entries serve as the core mechanism in SDN for directing traffic, establishing network paths, and implementing network policies. To effectively manage and analyze the flow tables, a unified representation model becomes essential. In this phase, SAFLA utilizes graph as the representation model to facilitate subsequent analysis and processing. 

The extraction of entities from network configuration rules is an essential step for grouping similar intents. This process not only enables a compact representation of these rules but also facilitates a bottom-up extraction of high-level intents. The procedure takes as its input flow tables from switches across the entire network. 

A flow table contains multiple flow entries, each defining the rules for processing specific traffic. Every flow entry comprises certain key fields. As depicted in Fig. \ref{flowentry}, this entry is designed for packets originating from IP 192.168.1.10 sent over TCP to the IP range 10.0.0.1/24 on port 443 with specific VLAN and MAC details. When this entry is matches, the switch modifies the packet's VLAN and destination MAC, forwards it to Port 6, and sends a copy to the controller. This entry has a high priority of 1500, has matched 3450 packets so far, and will auto-expire under certain conditions. To raise the abstraction level of a flow entry, we define each flow entry as: ${\rm{Entity}} = {\rm{\{ MatchFields}},{\rm{Priority}},{\rm{Acttion}},{\rm{Counters}},{\rm{Timeout\} }}$, where:
\begin{figure}
\centering
\includegraphics[width=3.3in]{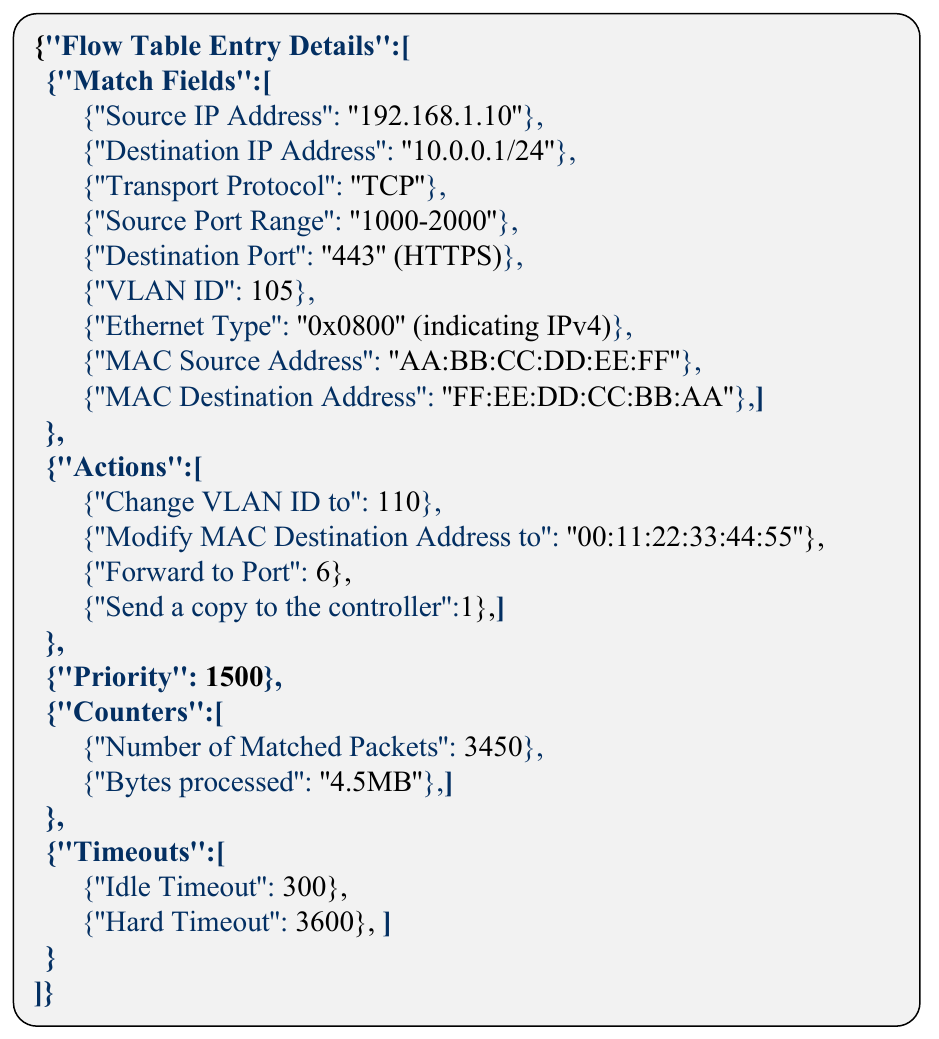}
\caption{Examples of flow entry details}
\label{flowentry}
\end{figure}
\begin{itemize}
\item \textbf{MatchFields:} determine which packets match the flow entry. Generally, match fields include but are not limited to the source IP address, destination IP address, source port number, destination port number, transport protocol, input port, Ethernet type, VLAN identifier, MAC address, etc. 

\item \textbf{Priority:} When multiple entries match a packet, this field determines which entry should be applied. Entries with a higher priority are considered first.
\item \textbf{Actions:} define the operations that should be executed on packets matching this entry. Common actions include forwarding to a specific port, Dropping the packet, Sending to the controller, etc. 
\item \textbf{Counters:} These track the number of packets matching the flow entry. 
\item \textbf{Timeout:} Some flow entries are temporary and might have a defined lifetime. When the timeout is reached, the entry can be automatically removed from the flow table.
\end{itemize}

\par

\begin{figure}
\centering
\includegraphics[width=3.0in]{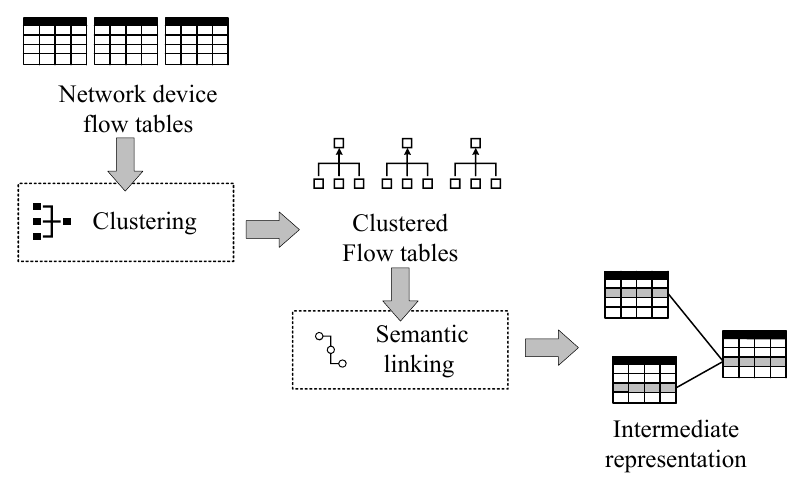}
\caption{Flow tables synthesis.}
\label{tablesythesis}
\end{figure}

\textbf{Clustering:}
We convert the flow tables into a preliminary representation that is structurally consistent. As depicted in Fig. \ref{tablesythesis}, we group the flow tables in each switch according to their endpoint groups. To be specific, flow entries with differing source and destination must be classified into distinct intents, whereas flow entries with identical addresses should be grouped to the same intent, which is denoted as ${S_i}$. Therefore, our grouping process aims to categorize flow entries according to their address information in the match fields. The function that defines the grouping of related flow entries within a switch is thus given by $$g:M \times N \to S$$
where ${S_i} \in {2^S}$. $M$ and $N$ are subset of $E$, representing flow table entries taken from each switch.  Formally, $\forall M,N \subset E$, if $M.addr \equiv N.addr$, then we group these two entries by $g(M,N)$. By taking this step, we can effectively reduce the redundancy of the flow table, thereby decreasing the total set of flow entries and consequently simplifying the complexity of the intent extraction process. 

\textbf{Semantic Linking:} It is important to highlight that a single forwarding intent typically spans configurations across several switches in SDN. Due to the semantically related nature of forwarding flow entries across these various switches, the amalgamation of grouped flow entries is imperative. We refer to this process as semantic linking. As a result, in the following phase, we interconnect the flow entries from all switches throughout the network, constructing a unified model that closely approximates the actual user's intent. Therefore, the process of semantic linking is given by $${f_1}:{S_1} \times {S_2} \times  \cdots  \times {S_n} \to {\rm Z}$$ where ${S_i}$ and ${\rm Z}$ are the clustered flow entries and the intermediate representations. Formally, $\forall {S_1},{S_2},...,{S_n} \in {2^S}$, if ${S_1}.addr \equiv {S_2}.addr \equiv ,..., \equiv {S_n}.addr$, then we link them by $f({S_1},{S_2},...,{S_n})$. Thus, each ${Z_i} \in {2^{\rm Z}}$ can be regarded as a long chain that collectively constitutes a singular intent. The significance of the linking function becomes clear in the context of associated intents that operate within the same domain. This approach guarantees that potential redundancies are prevented across different intermediate representations. In the following step, we will incorporate topological information from the NSKG to ascertain the correctness of these intermediate representations.

\begin{figure}
\centering
\includegraphics[width=2.1in]{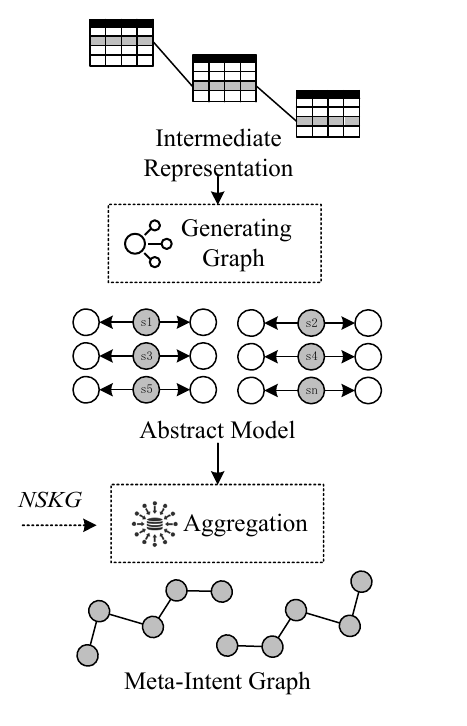}
\caption{The process of intent semantic fusion.}
\label{semanticfusion}
\end{figure}

%

\subsubsection{Semantic Fusion}

As illustrated in Fig. \ref{semanticfusion}, the semantic fusion process within our framework is meticulously designed to convert intermediate representations back into a comprehensible and high-level set of intents, leveraging the power of network state knowledge graphs for verification. This process unfolds in several structured steps as outlined below:

\textbf{Abstract Model:} This step involves mapping the previously obtained intermediate representations into a knowledge graph format, referred to as the abstract model. In this model, nodes represent switches, encapsulating detailed switch information, while edges denote the semantic linking between these nodes. This stage sets the foundation for transforming the raw data into a structured format that highlights high-level network functions and relationships.

\textbf{Aggregation:} Utilizing the Network Semantic Knowledge Graph (NSKG), this phase subjects each intermediate representation to rigorous scrutiny, focusing primarily on verifying the connectivity of switches within the actual network topology. This step, known as aggregation, is pivotal in ensuring that the abstracted representations reflect feasible network configurations, thereby validating the structural integrity of the modeled intents. It ensures that the abstract models are not only theoretically valid but also practically applicable within the current network state.

\textbf{Meta-Intent Graph:} Upon successful validation through NSKG, the abstract models are transformed into what we term as meta-intent graphs. These graphs not only precisely delineate high-level intents but also detail the connectivity paths through the network's switch nodes.

Through these steps, the semantic fusion phase delineates a robust methodology for back-translating intermediate representations into high-level network intents, grounded in the actual network topology and operational semantics. This approach not only enhances the precision of intent extraction but also enriches the network's semantic understanding, laying a solid foundation for subsequent intent consistency verification. This methodical process ensures that the derived intents are not only aligned with the network's current state but also provide a reliable basis for further analysis and optimization, thereby facilitating more informed and effective network management strategies.

\subsubsection{Consistency Check}
In this step, we systematically compare the endpoint group information extracted from each meta-intent graph with the intents stored in the intent repository. This comparison is performed by extracting the endpoint group information from the meta-intent graphs as tuples and matching them against the intents defined in the intent repository through a mathematical process. The core of this process involves defining two sets: ${\rm{G}} = {\rm{\{ }}{{\rm{g}}_1},{g_2},...,{g_n}{\rm{\} }}$ representing the set of all endpoint group information tuples in the meta-intent graphs, and ${\rm{I}} = {\rm{\{ }}{i_1},{i_2},...,{i_k}{\rm{\} }}$, representing the set of all intents in the intent repository. The goal is to find a bijective function ${f_2}:\rm{G} \to \rm{I}$ that matches each tuple ${g_n}$ from $\rm{G}$ to an intent ${i_j}$ in $\rm{I}$, indicating a one-to-one correspondence and thus semantic consistency between the extracted intents and the user's intents.

If a tuple ${g_i}$ in $\rm{G}$ cannot find a corresponding intent ${i_j}$ in $\rm{I}$, it suggests that the particular intent has become obsolete and needs to be reinstalled. Conversely, if there exists an intent ${i_j}$ in $\rm{I}$ without a matching tuple in $\rm{G}$, it signals an inconsistency, indicating that the configuration semantics and intent semantics are misaligned, necessitating further analysis to resolve the discrepancy.

\subsubsection{Repair Policy Reasoning}












We introduce two methods to effectively address semantic inconsistencies within the network's operational framework.

For scenarios where elements in G cannot be matched with tuples in I, this discrepancy is often due to structural changes in the network topology. Such changes might result from network upgrades or reconfigurations, potentially rendering certain intents obsolete or misaligned with the network's current configuration. To tackle this issue, our strategy includes a comprehensive review of the intents stored in the intent repository. We identify those impacted by the recent topology changes and proceed to reinstall them, ensuring they accurately represent the updated network layout. This procedure guarantees that the operational state of the network continues to reflect its strategic goals.

On the other hand, when elements in I cannot find corresponding tuples in G, it may signal unauthorized interventions, notably the injection of malicious flow entries by unauthorized applications. This scenario compromises both the network's intended functionality and its security. To counteract this, our framework leverages a combined top-down and bottom-up consistency check to implement a robust detection mechanism for such irregularities. Upon identifying these discrepancies, the framework purges the extraneous ${g_i}$ to maintain semantic consistency across the network.
  
By employing a methodical approach that combines top-down and bottom-up strategies within an INDs framework, we ensure the effective management of intents throughout their lifecycle. This holistic methodology not only facilitates the swift resolution of semantic inconsistencies but also enhances the network's security and resilience against unauthorized modifications, maintaining the integrity and alignment of network operations with the organizational objectives.

\section{Performance Evaluation}

This section evaluates the performance of the SAFLA framework across various application scenarios and network configurations. Initially, we conduct a case study to demonstrate SAFLA's effectiveness in a corporate network setting, focusing on its capability to ensure semantic consistency and counteract potential hijacking attack. Subsequently, we examine the efficiency of our proposed bottom-up approach for high-level intent extraction, which is essential for the framework's functionality. Furthermore, we have benchmarked SAFLA with the ONOS's Intent Framework \cite{OIF}, comparing their abilities to maintain intent viability during hijacking attacks in runtime. Lastly, we have conducted stress tests on the SAFLA framework to evaluate its scalability on large network with an increase in the number of intents and switches.

\subsection{Evaluation Environment}
To thoroughly access the SAFLA framework, we employ a network topology ranging from 1 to 400 nodes. This topology was realized through the use of Containernet \cite{Containernet}. Containernet is a versatile network emulation tool, allows for the creation and manipulation of network topologies in a controlled environment. This tool was instrumental in constructing the varied network topologies necessary for our tests. The experimental setup was hosted on a robust Ubuntu desktop equipped with 64GB RAM and an Intel(R) Core(TM) i7-9700K CPU operating at 3.60GHz. The desktop was running Ubuntu 20.04.1, with resources efficiently allocated among Containernet (to handle hosts and Open vSwitch), the SDN Controller (ONOS version 2.7.0), and the SAFLA framework itself. This configuration ensured a reliable and realistic environment for evaluating the performance and scalability of SAFLA under various network conditions.

\subsection{Corporate Network Scenario}
In this section, we present a case study of a small company that illustrates the applicability of our solution in maintaining the intent life cycle, demonstrating its effectiveness and technical feasibility. The goal is to ensure that network behavior semantically aligns with its declared intent throughout its full lifecycle, thereby avoiding potential disruptions during the operational phase. 

\begin{figure}[t]
\centering
\includegraphics[width=3.5in]{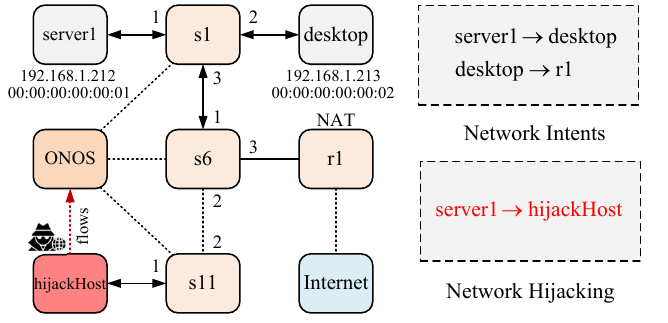}
\caption{A case study of a small corporate network with SDN switches}
\label{Visio-usecase1}
\end{figure}

As shown in Fig. \ref{Visio-usecase1}, the case study outlines a network with two primary intents. However, it also highlights a potential threat: a rerouted flow table with a higher priority between 'desktop' and 'server1', often overlooked by administrators due to its stealthy nature and the complexity of analyzing the underlying flow tables. Conventionally, network administrators are required to engage in continuous monitoring, which includes checking anomaly logs and observing network behavior. This task encompasses regular reviews of network performance metrics, timely adjustments to configurations as necessary, and comprehensive testing to ensure each intent functions as intended throughout its lifecycle. However, this approach to network management is often challenged by the reactive nature of traditional network management practices. Many business disruptions related to network intents only become apparent after they have caused significant disruptions. Consequently, administrators might only become aware of these issues after they have led to notable performance degradation or even critical system disruptions. This delayed detection and response underscores the need for a shift towards a more proactive framework that maintains the semantic consistency between high-level intent and low-level configurations. 


\begin{figure}[!t]
\centering
\includegraphics[width=3.0in]{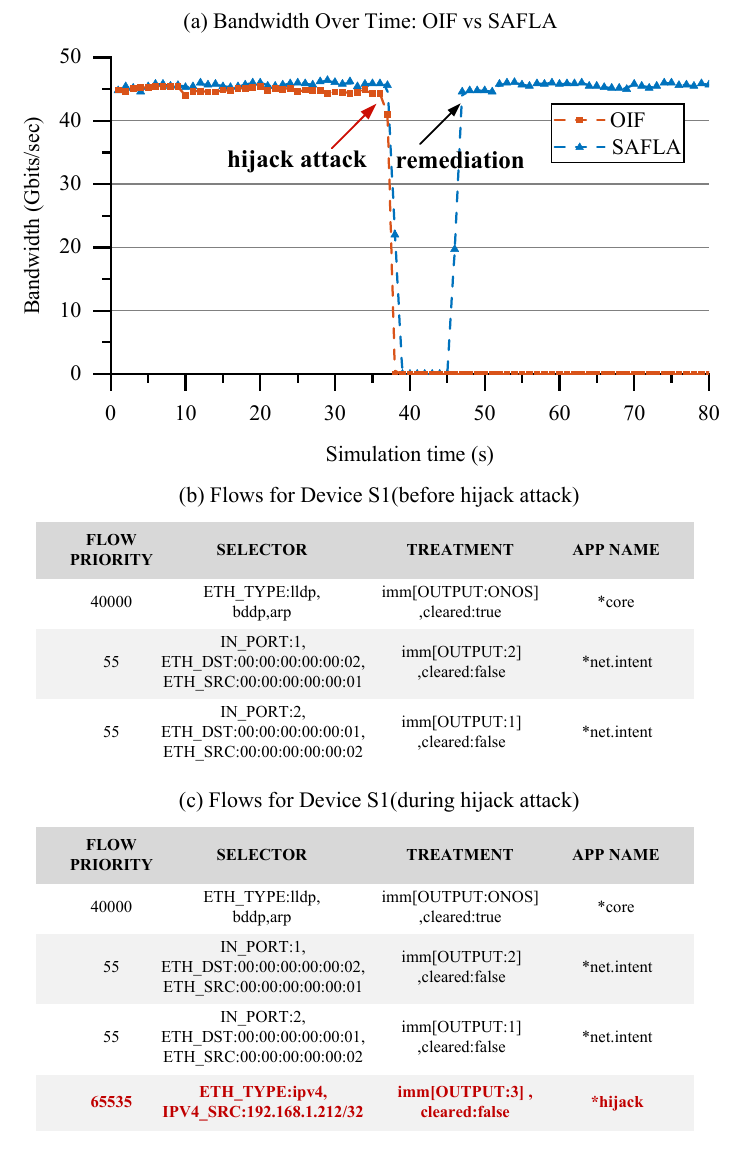}
\caption{This analysis showcases the performance of the proposed framework in ensuing intent full life assurance. As the intensity of attacks escalates, including increased disrupted topology nodes and malicious flow entries, the framework consistently demonstrates a robust intent survival rate.}
\label{bandwidth}
\end{figure}

\begin{table*}
    \renewcommand{\arraystretch}{1.3} 
    \caption{Comparison of the extraction times (in seconds) of FWS SCRIBE and the proposed SAFLA.}
    \label{comparison_intents_time}
    \centering
    \begin{tabular}{lccccccccccccccc}
    \toprule
    \multirow{2}{*}{\textbf{Methods}} &\multirow{2}{*}{\textit{Info.}} & \multirow{2}{*}{\makecell{\textit{Holistic} \\ \textit{Analysis}}}& \multirow{2}{*}{\makecell{\textit{Standalone} \\ \textit{Capability}}} &\multicolumn{10}{c}{\textbf{Number of Intents vs. Processing Time }} \\
    \cmidrule(lr){5-14}
    & & &&  \textbf{10} & \textbf{20}& \textbf{30} & \textbf{40} & \textbf{50}& \textbf{60}& \textbf{70} & \textbf{80}& \textbf{90}& \textbf{100} \\
    \midrule
    SCRIBE\cite{scribe} & iptables         &\xmark &\xmark& 0.2071 & 0.4333 & 0.8037 & 1.3124 & 2.0986& 3.0047 & 4.1298 & 5.4272& 7.2575&8.6499\\
    FWS \cite{fws}           &iptables&\xmark &\cmark& 0.2064 & 0.4336 & 0.8028 & 1.3136 & 2.0972& 3.0031 & 4.1274 & 5.4251& 7.2552&8.6457\\
    \textbf{SAFLA}&flowtables &\cmark&\cmark & \textbf{0.0001}& \textbf{0.0002} & \textbf{0.0004} & \textbf{0.0005} & \textbf{0.00006} & \textbf{0.0007}& \textbf{0.0009} & \textbf{0.0013} & \textbf{0.0015}&\textbf{0.0015} \\
    \bottomrule
    \end{tabular}
\end{table*}






In our proposed SAFLA, once an intent is successfully translated and implemented in hardware, it initiates a continuous assurance loop that persists throughout the intent's lifecycle. This involves periodically monitoring intents extracted from lower-level configurations, ensuring they align with the original user intent. Maintaining this alignment from the bottom up is crucial to preserve the semantic integrity of the intent and guarantee semantic consistency. In cases of discrepancies, an intent remediation process is initiated. If a rectification policy is applicable, this intent will then adopt a top-down approach again to repair/remove malicious flows to maintain the intent's semantic integrity.

To validate the effectiveness of our proposed SAFLA approach, we conducted a comparative experimental analysis against the ONOS's intent framework that employs a strictly top-down design. As depicted in Fig. \ref{bandwidth}(a), we initiated an intent that maintains a desktop's request for resources from server1, with the flow table illustrated in Fig. \ref{bandwidth}(b). Initially, from 0 to 37 seconds, the services under both SAFLA and ONOS's intent frameworks were operational, reaching a transmission rate of approximately 45 Gbit/sec. However, at the 37.5-second mark, a simulated hijack attack was executed, diverting all traffic intended for the desktop to the hijackHost, detailed in the flow table in Fig.\ref{bandwidth} (c). The immediate consequence was a cessation of service, with the transmission rate plummeting to zero in both approaches. Notably, within approximately 5 seconds, the services utilizing the SAFLA framework demonstrated a robust recovery. This resilience is made possible by continuously reconciling the lower-level configurations with the original user intent through an integrated assurance loop. Upon detecting discrepancies, SAFLA promptly enacted a rectification policy, purging the errant flow tables and thus resuming the intended services.

\subsection{Intent Extraction Performance}

To assess the feasibility of our framework, we explored its applicability in the context of SDN routing. Our evaluation hinged on two critical factors: (\textit{i}) the capability of our framework to efficiently extract and generate remediation intent; and (\textit{ii}) the resilience demonstrated by the intent survival rate, particularly when intents faced with disruptions due to switch breakdown and malicious flow injection.

\subsubsection{Single Device Analysis}

To validate the performance of our proposed framework, we conduct a comparative analysis focusing on the time efficiency of intent extraction. In this analysis, we compared our framework with FWS and SCRIBE, which are both network configuration analyzing tools that automatically interpret network behavior in higher-level representations (intents). Our comparison involves extracting high-level intents, varying in quantity from 10 to 100. The results of this evaluation are presented in Table \ref{comparison_intents_time}, where the best outcomes are highlighted in bold. The 'Info' column in the tables provides details on the input for each method: 'iptables', a widely used packet filtering tool in Linux operating systems that functions as a software firewall, and 'Flowtables', a high-performance, programmable forwarding engine in software-defined networking environments, also capable of acting as a software firewall.

For the single-node intent extraction experiment, we employ Containernet to emulate a network environment that mirrors the one under iptables. This emulation is achieved by employing SDN switches within Containernet to simulate firewall functionalities, thereby allowing us to ensure uniformity in network capabilities across all three tested methods. To ensure this, we focus on accessibility intents, which permit the forwarding of specific IPs or protocols. Meanwhile, both FWS and SCRIBE utilize iptables' filter rules, creating a level basis for comparing these different intent extraction methods. This uniformity in network setup and application of rules ensures that all methods are evaluated under consistent conditions, particularly at the configuration level, thereby facilitating a fair comparison.

\begin{figure}[!t]
\centering
\includegraphics[width=3.0in]{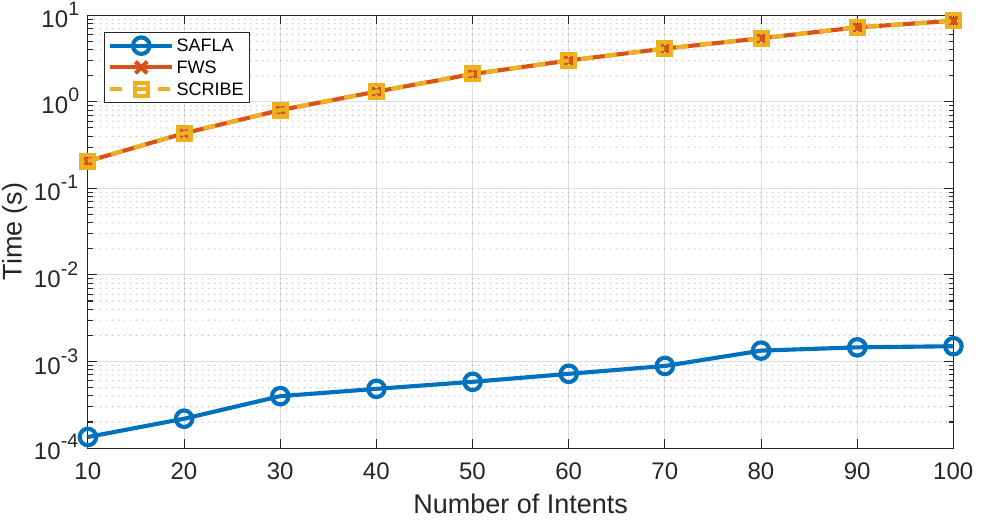}
\caption{A comparison of the intent extraction time across SCRIBE, FWS, and SAFLA, demonstrating that SAFLA significantly outperforms the other two methods. At the same number of intents, SAFLA's extraction time is up to four orders of magnitude faster than both SCRIBE and FWS.}
\label{time_compare}
\end{figure}

As highlighted in Table \ref{comparison_intents_time}, our experiment reveals that the proposed OFIAL outperforms FWS and SCRIBE in terms of time efficiency. Figure \ref{time_compare} illustrates our framework's superior efficiency over a range of 10 to 100 intents, consistently exceeding others at least by an order of magnitude. This disparity in efficiency becomes even more pronounced at the 100-intent benchmark, where our method is observed to be three orders of magnitude faster in time efficiency compared to both SCRIBE and FWS.

These results unequivocally highlight the superior performance of our framework in single-node scenarios, indicating its potential for more effective application in large-scale network intent consistency assurance, compared with the other methods examined. Furthermore, our framework has the advantage of being able to analyze global network intents in scenarios involving more than two switches. In the following subsection, we evaluate the SAFLA framework's capability in extracting intents at a large-scale network level.

\subsubsection{Multi-device analysis}
We access the SAFLA framework's performance in extracting intent within a network that comprises 100 intents. Our experimental setup explores a range of network sizes by progressively increasing the number of switch nodes. Starting from a single node, the network expands to include 50 (10x5), 100 (10x10), 200 (10x20), and 400 (10x40) switch nodes, all interconnected in a mesh topology to ensure maximal connectivity. Furthermore, we arrange 20 hosts as the endpoints for these intents. These hosts are placed at the network's polar ends, which allows for the flow entries to be evenly installed across the switches in the network.

\begin{figure}[!t]
\centering
\includegraphics[width=3.0in]{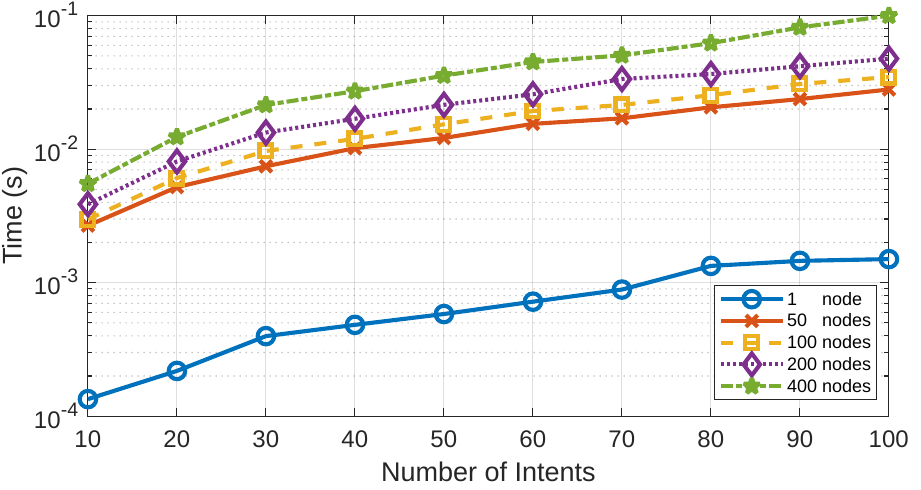}
\caption{A comparison of intent extraction times across various switch node sizes reveals sub-linear scaling with increasing intent counts. Remarkably, with a network of 400 nodes processing 100 intents, the extraction time is maintained at a mere 0.1 seconds.}
\label{topo_size}
\end{figure}

As depicted in Fig. \ref{topo_size}, we observe a stable increase in the time required for intent extraction as the number of nodes grows. Crucially, there is no abrupt surge in the time needed, despite the network's increasing complexity. This steadiness underscores the framework's scalability and robustness, illustrating its ability to handle complex network configurations without a loss in performance efficiency. Moreover, the SAFLA framework particularly excels when managing a network with 100 intents across 400 switch nodes, achieving intent extraction in a swift 0.1 seconds. This impressive result not only demonstrates its capability for real-time efficiency but also establishes its potential to facilitate full lifecycle intent assurance at runtime.


\subsection{Runtime Performance}
In this section, we evaluate the SAFLA framework's ability to maintain intent's full lifecycle during runtime. Our assessment focuses on the framework's handling of intent disruptions, particularly those arising from topology changes affecting connectivity intents and hijack attacks impacting existing intents. Furthermore, we test the framework's assurance performance at runtime with different topology sizes and number of intents, highlighting its operational robustness and adaptability in real-world applications.

\subsubsection{Intent Assurance}
In our evaluation, we sought to assess the efficacy of our proposed SAFLA framework in managing the lifecycle of network intents under conditions of varying network topology completeness, from 40\% to 100\%. This range represents the integrity of the network topology, with 100\% completeness indicating a fully intact topology and 40\% completeness indicating that 60\% of switch nodes are non-functional, simulating the effect of partial to complete switch failures on network operations. We incorporated ten critical business intents into a network simulation to ensure operational continuity. The performance of our framework was benchmarked against the ONOS's intent framework which employs a standard primary backup recovery algorithm, by measuring the intent survival rate in scenarios of fixed topology. Our findings illustrated that our framework consistently surpassed the standard backup algorithm across varying levels of topology completeness, significantly enhancing robustness and the rate of intent recovery even in conditions where network integrity was compromised.

Specifically, our framework demonstrated a marked superiority when the network's topology completeness was at 80\%, doubling the performance of the backup algorithm, and further distinguished itself at 90\% completeness, where our intent survival rate exceeded that of the backup algorithm by 36\%. These results underline our system's exceptional efficiency in achieving near-total remediation, particularly under adverse conditions, and confirm its effectiveness in restoring business intents. As the completeness of the network topology improved to and beyond 80\%, both frameworks showed an improved intent survival rate. Notably, our framework almost achieved full remediation at 90\% completeness. The ONOS's intent framework with the Primary backup algorithm, however, showed comparable effectiveness only when the network's integrity was near-perfect or at full capacity.

This experiment reveals that our proposed SAFLA framework is inherently more resilient to significant network impairments, providing a more reliable solution for intent survival rate. Such resilience is crucial in maintaining uninterrupted services, particularly in situations where network stability cannot be guaranteed. The findings underscore the potential of our framework to adapt dynamically to varying levels of network disruptions, ensuring service continuity even in the most unpredictable of scenarios.

\begin{figure}[!t]
\centering
\includegraphics[width=3.0in]{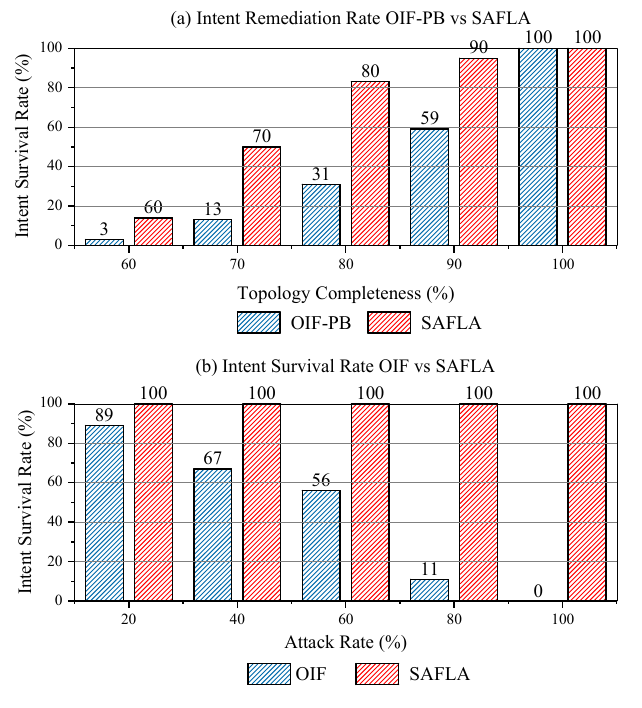}
\caption{This analysis showcases the performance of the proposed framework in ensuing intent full life assurance. As the intensity of attacks escalates, including increased disrupted topology nodes and malicious flow entries, the framework consistently demonstrates a robust intent survival rate.}
\label{assurance}
\end{figure}

To further complement this experiment, we also evaluated the resilience against simulated hijacking attacks, which compromise deployed intents by targeting the network's flow tables. To control the intensity of the attacks, we introduced varying numbers of malicious flow entries into the data plane, noting that higher attack intensities result in more disrupted intents. Our methodology encompasses a thorough analysis of flow entries in each SDN switch to extract and assess higher-level intents. By comparing these with the user's original intents stored in the intents repository, our framework ensures the integrity and consistency of user intents across their lifecycle by either reinstalling or purging compromised flow entries.  This approach effectively bridges the semantic gap between actual configurations and the declarative high-level intents. In comparison, we assessed ONOS's intent framework to maintain intent consistency under similar adverse conditions. For the experimental configuration, we established a network topology using Containernet, integrating a switch connected to 11 hosts, over which we installed 9 unique intents for assurance test. As depicted in Fig. \ref{assurance}(b), our framework significantly surpasses the ONOS intent framework in maintaining intent consistency across various attack intensities. Notably, our framework maintains a nearly constant intent consistency rate, unaffected by the severity of the attack. Even with the attack intensity reaching up to 100\%, our framework's ability to preserve intent integrity remains staunchly robust. In contrast, the ONOS's intent framework exhibits a pronounced decline in intent consistency with increasing attack rates. This decline is attributed to ONOS's intent framework's top-down intent management methodology, which relies on predefined states to accommodate potential variations throughout the intent's lifecycle—from compilation to post-installation phases. At a 100\% attack rate, the ONOS framework's capacity to ensure intent consistency is virtually nullified, highlighting a critical susceptibility to network threats. This contrast underscores our framework's superior detection and recovery capabilities, demonstrating its strength in maintaining intent consistency even under severe network threats. The experiment emphasizes the importance of a holistic approach that encompasses the entire lifecycle of network intents. It also confirms the effectiveness of our proposed solution in safeguarding the network's intent consistency against network dynamics, offering a robust mechanism for intent assurance.

\subsubsection{Assurance Stalibility}
\begin{table}[h!]
    \renewcommand{\arraystretch}{1.3} 
    \caption{Impact of the number of intents on hijack attack detection/remediation latency.}
    \label{comparison_intents}
    \centering
    \begin{tabular}{lcc}
    \toprule
    \textbf{No. Intents} & \textbf{Recover Time} & \textbf{Time Increment} \\
    \midrule
    20           &2.1925 s &-\\
    40           &2.2056 s &+0.597\% \\
    60           &2.2034 s  &-0.099\%\\
    80           &2.2372 s &+1.534\% \\
    100           &2.2851 s &+2.141\%\\
    \bottomrule
    \end{tabular}
\end{table}

\begin{table}[h!]
    \renewcommand{\arraystretch}{1.3} 
    \caption{Impact of the number of switches on hijack attack detection/remediation latency.}
    \label{comparison_nodes}
    \centering
    \begin{tabular}{lcc}
    \toprule
    \textbf{No. Switches} & \textbf{Recover Time} & \textbf{Time Increment} \\
    \midrule
     1           &0.0365 s &-\\
    50           &0.7363 s &+1917.26\%\\
    100           &2.2034 s &+199.253\% \\
    200           &8.1112 s &+268.122\%\\
    400           &33.8697 s &+317.567\%\\
    \bottomrule
    \end{tabular}
\end{table}

To assess SAFLA's real-time performance with a varied number of intents and network sizes, we explored its capability to ensure intent consistency during runtime.

Our initial setup involved a fixed network configuration with 100 SDN switches. Table \ref{comparison_intents} reveals that as we scaled the number of intents from 20 to 100, the time SAFLA needed to recover from inconsistencies showed a slight rise from 2.1925 to 2.2851 seconds. The time increment peaked at just over 2\% for the largest set of intents. This finding illustrates that SAFLA can efficiently handle an increase in intents without significantly longer recovery times, demonstrating high analysis efficiency in stable network topologies.

In contrast, when we fixed the number of intents at 60—due to the performance constraints of the ONOS framework—and varied the network size, we observed a different pattern. As depicted in Table \ref{comparison_nodes}, the recovery time increased markedly from 0.0365 seconds with one switch to 33.8697 seconds with 400 switches. This substantial rise is mainly due to the increased time needed to export and process the flow tables from a greater number of switches. However, this challenge could potentially be mitigated in the future by parallel processing of flow table exports, which would shorten the overall time required for larger topologies.

To sum up, the SAFLA framework's scalability is evident. It effectively ensures intent assurance quickly across different numbers of intents and sizes of network topologies. While the time increase is more noticeable in larger topologies, it remains within a feasible range, confirming SAFLA's capability for large-scale network application.

\section{Conclusion}

In this paper, we have presented SAFLA, a novel semantic-aware framework that assures the full lifecycle of intents within intent-driven networks (IDNs). SAFLA innovatively blends bottom-up insights with the conventional top-down approach to effectively address disruptions and bridge the semantic gap between the network's high-level intents and their executable configurations. Through meticulous case studies, we developed a method to rectify semantic inconsistencies, ensuring a robust alignment between the network’s operational state and its intended objectives. Experiments were conducted to evaluate SAFLA's performance in extracting intents and its operational feasibility in real-time environments. The results affirm that SAFLA excels in detecting and resolving semantic inconsistencies during runtime, significantly enhancing the effectiveness of intent management across its entire lifecycle. These findings underscore the pivotal role of SAFLA in advancing the capabilities of IDNs, ensuring that they can adapt dynamically to changes while maintaining high levels of accuracy and reliability in intent execution.

\vspace{11pt}
\vfill
\end{document}